\documentclass[aps,prl,twocolumn,showpacs,superscriptaddress,amsmath,amssymb]{revtex4-1}
\usepackage{graphicx}
\usepackage{color}

\definecolor{Red}{rgb}{1.0,0.0,0.0}

\begin{document}

\title{Enhanced Flow in Small-World Networks}

\author{Cl\'audio L. N. Oliveira}%
\email{lucas@fisica.ufc.br}%
\affiliation{Departamento de F\'isica, Universidade Federal do
  Cear\'a, 60451-970 Fortaleza, Cear\'a, Brazil}%
\affiliation{Department of Biomedical Engineering, Boston University,
  Boston, MA 02215}%

\author{Pablo A. Morais}
\email{pablo@fisica.ufc.br}
\affiliation{Departamento de F\'isica, Universidade Federal do
  Cear\'a, 60451-970 Fortaleza, Cear\'a, Brazil}%

\author{Andr\'e A. Moreira}
\email{auto@fisica.ufc.br}
\affiliation{Departamento de F\'isica, Universidade Federal do
  Cear\'a, 60451-970 Fortaleza, Cear\'a, Brazil}%

\author{Jos\'e S. Andrade Jr.}
\email{soares@fisica.ufc.br}
\affiliation{Departamento de F\'isica, Universidade Federal do
  Cear\'a, 60451-970 Fortaleza, Cear\'a, Brazil}%


\begin{abstract}
  The small-world property is known to have a profound effect on the
  navigation efficiency of complex networks [J. M. Kleinberg, Nature
  406, 845 (2000)]. Accordingly, the proper addition of shortcuts to a
  regular substrate can lead to the formation of a highly efficient
  structure for information propagation. Here we show that enhanced
  flow properties can also be observed in these complex topologies.
  Precisely, our model is a network built from an underlying regular
  lattice over which long-range connections are randomly added
  according to the probability, $P_{ij}\sim r_{ij}^{-\alpha}$, where
  $r_{ij}$ is the Manhattan distance between nodes $i$ and $j$, and
  the exponent $\alpha$ is a controlling parameter. The mean two-point
  global conductance of the system is computed by considering that
  each link has a local conductance given by $g_{ij}\propto
  r_{ij}^{-\delta}$, where $\delta$ determines the extent of the
  geographical limitations (costs) on the long-range connections. Our
  results show that the best flow conditions are obtained for
  $\delta=0$ with $\alpha=0$, while for $\delta \gg 1$ the overall
  conductance always increases with $\alpha$. For $\delta\approx 1$,
  $\alpha=d$ becomes the optimal exponent, where $d$ is the
  topological dimension of the substrate. Interestingly, this exponent
  is identical to the one obtained for optimal navigation in
  small-world networks using decentralized algorithms.
\end{abstract}

\maketitle

The Laplacian matrix operator is a general description for systems
presenting two essential properties: ($i$) they obey local
conservation of some load, and ($ii$) their currents of load are
linearly dependent on some field~\cite{Bollobas2012}.  Since such
conditions are very often observed, this operator can be applied to
several different problems in Physics, including
diffusion~\cite{Monasson1999,Andrade1997}, wave
propagation~\cite{Conti2004}, solving the Schr\"odinger equation in
arbitrary graphs~\cite{Beylkin2001}, dielectric
breakdown~\cite{Kahng1988}, brittle
fracture~\cite{Moreira2012,Oliveira2012}, Darcy's flow~\cite{Lee1999}
and classical electrical
transport~\cite{Kirkpatrick1971,Cheainov2007}, among others. In these
problems, one is frequently interested in the stationary state, where
currents in each edge of a given network can be determined using local
conservation laws.  In the field of complex networks, in particular,
the Laplacian operator has been employed as a conceptual approach for
determining the nature of the community structure in the
networks~\cite{Newman2004}, in the context of network
synchronization~\cite{Korniss2006}, as well as to study network
flow~\cite{Lopez2005,Novotny2007,Lee2006,Helbing2006,Danila2007,Andrade2011}.

Given a regular network as an underlying substrate, it has been shown
that the addition of a small set of random long-range links can
greatly reduce the shortest paths among its sites. In particular, if
the average shortest path $\ell_{s}$ grows slowly with the network
size $N$, typically when $\ell_{s}\sim\log(N)$, the network is called
a {\it small world}~\cite{Albert1999,Watts1998,Newman1999}. If one
considers the effect of constraining the allocation of the long-range
connections with a probability decaying with the distance, $P_{ij}\sim
r_{ij}^{-\alpha}$, results in an effective dimensionality for the
chemical distances that depends on the value of
$\alpha$~\cite{Moukarzel2002}. For the case in which the regular
underlying lattice is one-dimensional, the small-world behavior has
only been detected for $\alpha<2$, with $\ell_{s}$ reaching a minimum
at $\alpha_{opt}=0$~\cite{Moukarzel2002}.  The two-dimensional case
was also investigated~\cite{Kosmidis2008}, yielding similar results.

The situation is much more complex if one does not have the global
information of all the short-cuts present in the network. As a
consequence, the traveler does not have {\it a priori} knowledge of
the shortest path. Optimal navigation with local knowledge and the
presence of long-range links was studied by
Kleinberg~\cite{Kleinberg2000}.  Surprisingly, the small-world
features of the network can only be efficiently accessed if the
exponent is precisely set at $\alpha=d$, where $d$ is the topological
dimension of the substrate. It is then claimed that this condition is
optimal due to the presence of strong correlations between the
structure of the long-range connections and the underlying lattice,
leading to the formation of ``information gradients'' that allow the
traveler to find the target. Later, it was shown that, by imposing a
cost constraint to the long-range connections, results in
$\alpha_{opt}=d+1$, for both local and global knowledge
conditions~\cite{Li2010}.

A question that naturally arises from these navigation studies is how
efficient small-world networks are for transport phenomena that
typically obey local conservation laws. Here we show that enhanced
Laplacian flow properties can also be observed for networks built by
adding long-range connections to an underlying regular lattice, in the
same fashion as previously proposed for navigation through small-world
geometries~\cite{Kleinberg2000,Kosmidis2008,Li2010,Barthelemy2010,Gallos2012}.
Our model consists of $N$ nodes arranged on a circle, and connected to
their two nearest neighbors. Long-range connections~\cite{notsolong}
are added to the regular substrate by ensuring that each node $i$
receives a new link to a node $j$ randomly chosen among those $N-3$
remaining nodes according to the probability, $P_{ij}\sim
r_{ij}^{-\alpha}$, where $\alpha$ is an arbitrary exponent, and
$r_{ij}$ is the Manhattan distance, namely, the minimum number of
links separating nodes $i$ and $j$. The larger the parameter $\alpha$,
the shorter are the long-range links.  In fact, given the probability
distribution and the size constraints, for $\alpha<d$, the average
link length grows with the system size, $\langle{r}\rangle\sim{N}$,
while for $d<\alpha<d+1$, $\langle{r}\rangle\sim{N^{d+1-\alpha}}$, and
for $\alpha>d+1$, $\langle{r}\rangle\sim{N^0}$. In the limiting cases,
one may expect a logarithmic dependence, $\langle{r}\rangle\sim
N/\log(N)$ for $\alpha=d$, and $\langle{r}\rangle\sim \log(N)$ for
$\alpha=d+1$.

\begin{figure}[t]
  \includegraphics*[width=5.5cm]{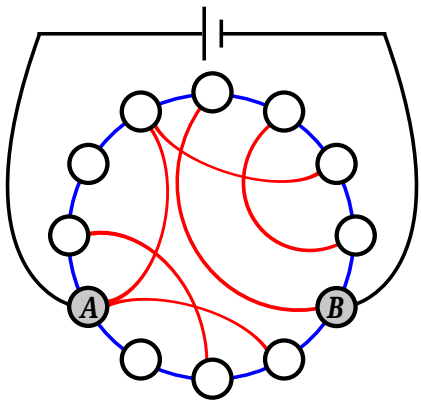}
  \caption{(Color online) Network built on an one-dimensional
    underlying lattice of nearest neighbors (short-range) connections
    (in blue) and long-range connections (in red). Each connection has
    a local conductance given by $g_{ij}$. A unity global current is
    applied through the system between nodes $A$ and $B$ so that a
    global conductance, $G_{AB}$, can be computed. We obtain the
    average conductance, $\langle G\rangle$, by calculating $G_{AB}$
    over different pairs of sites $AB$ and several network
    realizations.}
  \label{f.1D}
\end{figure}

Once the network is built, we associate each link to an Ohmic resistor
and a unitary global current is induced between a pair of sites $A$
and $B$ in the system (see Fig.~\ref{f.1D}). In order to compute the
local potential $V_i$, we solve Kirchhoff's law at each site
$i$~\cite{HSL},
\begin{equation}
  \sum_j g_{ij}(V_i-V_j) = 0, \,\,\,\, i=1,...,N,\nonumber
\end{equation}
with the summation running over all sites connected to $i$, and
$g_{ij}$ being the link conductance between $i$ and $j$. The inlet and
outlet currents are also considered in the calculation of the local
potentials at sites $A$ and $B$, respectively. The global conductance
for a given realization, which depends on the chosen pair of sites $A$
and $B$, is computed as $G\equiv 1/\Delta V$, where $\Delta
V=V_A-V_B$, so that a mean global conductance $\langle G\rangle$,
between any two sites of the network, is then obtained by averaging
over different pairs of sites~\cite{footnote} and different
realizations of the network.

Here we assume that the conductance of each link depends on the
distance between its ends in the form,
\begin{equation}
  g_{ij}=r_{ij}^{-\delta},\nonumber
\end{equation}
where the exponent $\delta$ gauges the way long-range connections
impact the flow, as $\delta$ increases the contribution of longer
connections to flow is attenuated. We focus on two particular cases,
namely, $\delta=0$ and $1$. In the case of $\delta=1$, known as
Pouillet's law~\cite{Kipnis2009}, the conductance of a long-range
connection with Manhattan distance $r$ is equivalent to an effective
conductance of $r$ short-range links in series, hence one should
expect long-range links contributing to transport as much as
short-range links.  Accordingly, for $\delta<1$ ($\delta>1$),
preferential flow should occur through long-range (short-range) links.
\begin{figure}[t]
  \includegraphics*[width=8.3cm]{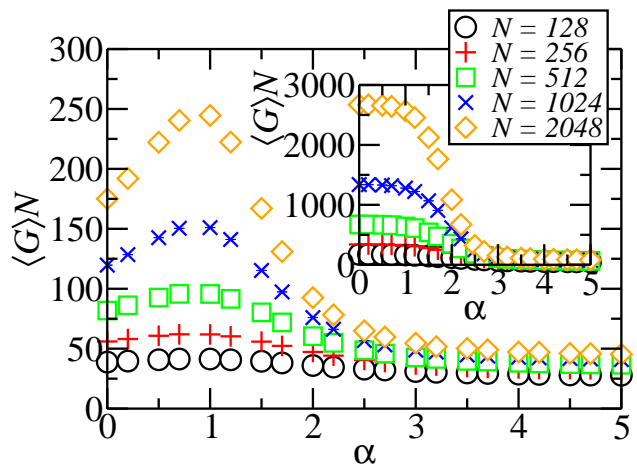}
  \caption{(Color online) Mean effective conductance versus the
    exponent $\alpha$ for $\delta=1$ (main graph) and $\delta=0$
    (inset), in one-dimensional substrates, and for different system
    sizes. The maximum conductance is obtained at $\alpha=1$, in the
    first case (local conductances inversely proportional to the
    length), and at $\alpha=0$, in the second case (conductances are
    independent of link length). The error bars are smaller than the
    symbols.}
  \label{cond1D}
\end{figure}

As shown in Fig.~\ref{cond1D} (inset), the network conductance
$\langle G\rangle$ decays monotonically with $\alpha$ for $\delta=0$,
regardless of the system size $N$, since the shortcut links in this
case participate very actively in the flow. The maximum conductance is
therefore obtained at $\alpha=0$. The main panel of Fig.~\ref{cond1D}
shows, however, that the average conductance calculated for $\delta=1$
behaves non-monotonically with the exponent $\alpha$, with a maximum
value observed at $\alpha=1$.

\begin{figure}[t]
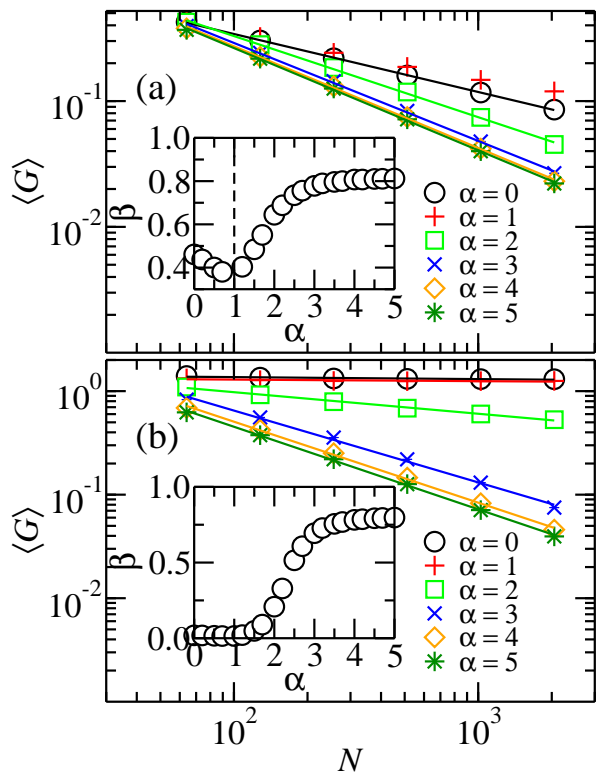

  \includegraphics*[width=7.8cm]{condxL_delta10.eps}
  \includegraphics*[width=7.8cm]{condxL_delta00.eps}
  \caption{(Color online) Dependence on size of the mean effective
    conductance for different values of $\alpha$. In (a) we show the
    case in which local conductances are inversely proportional to the
    length ($\delta=1$). The average conductances generally follow a
    power-law behavior, $\langle G\rangle\sim N^{-\beta}$, except for
    the optimal condition $\alpha=1$, where $\langle G\rangle$ behaves
    approximately as a power law of the logarithm of $N$, as depicted
    in Fig.~\ref{condxL-2}. The exponents resulting from the
    least-squares fitting to the data of the scaling functions are
    shown in the inset, except for the case $\alpha=1$ (dashed line).
    A non-monotonic behavior can be clearly observed. The same is
    shown in (b), but for $\delta=0$. In this case, the average
    conductance always obeys a power law. The inset shows that the
    exponent $\beta$ increases monotonically with $\alpha$.}
  \label{condxL}
\end{figure}

In Figs.~\ref{condxL}(a) and \ref{condxL}(b) we show the dependence of
the average network conductance on the network size $N$ for $\delta=1$
and $0$, respectively. The results in Fig.~\ref{condxL}(a) indicate
that, except for $\alpha=1$, a typical power-law behavior is observed,
$\langle G\rangle\sim N^{-\beta}$, with an exponent $\beta$ that
depends strongly on the parameter $\alpha$ controlling the length of
long-range connections.  In this case, the local conductance is
inversely proportional to the distance. As presented in the inset of
Fig.~\ref{condxL}(a), the exponent $\beta$ starts from $0.46$, at
$\alpha=0$, and falls to a minimum as $\alpha$ approaches unity. For
$\alpha>1$, the exponent $\beta$ again grows continuously. In the
limit of large values of $\alpha$, the added links always connect
close sites, $\langle r\rangle\sim N^{0}$.  As a consequence, the
system should behave as a regular lattice, $\langle G\rangle\sim
N/\log(N)$, regardless of the value of $\delta$.  In the case of
$\delta=0$, all links have identical resistances and a power-law
behavior is always observed, as shown in Fig.~\ref{condxL}(b). The
inset of Fig.~\ref{condxL}(b) shows that the exponent $\beta$ vanishes
for small values of $\alpha$, meaning that the average conductance
becomes independent of $N$.  However, as $\alpha$ increases, the
conductance exponent approaches unity, the analytical result for a
regular one-dimensional lattice.

\begin{figure}[t]
  \includegraphics*[width=8.3cm]{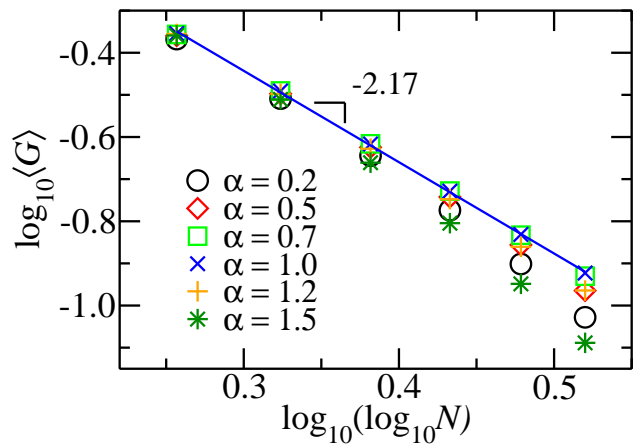}
  \caption{(Color online) For $\delta=1$ and $\alpha=1$, the global
    conductance follows a power-law of the logarithm of $N$, $\langle
    G\rangle\propto (\log_{10}N)^{-\gamma}$. The least-squares fitting
    to the data points for $\alpha=1$ gives an exponent $\gamma=2.17
    \pm 0.02$.}
  \label{condxL-2}
\end{figure}

Going back to the optimal flow condition, namely, $\delta=1$ and
$\alpha=1$, since logarithmic corrections are also present, an
exponent $\beta\rightarrow 0+$ is expected. To support this
conjecture, we show in Fig.~\ref{condxL-2} that the average
conductance in this particular case follows a power law of the
logarithm of $N$, $\langle G\rangle\propto (\log_{10}N)^{-\gamma}$,
with an exponent $\gamma=2.17 \pm 0.02$.

\begin{figure}[t]
  \includegraphics*[width=8.3cm]{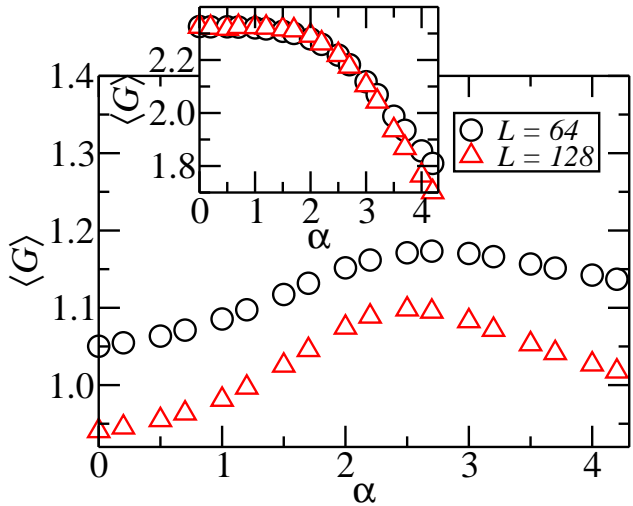}
  \caption{(Color online) Dependence on $\alpha$ of the average global
    conductance of networks built by adding long-range connections on
    two-dimensional regular lattices. For $\delta=0$ (main graph), the
    maximum conductance is obtained at $\alpha=0$.  In the case of
    $\delta=1$, our results show optimal exponents at $\alpha\approx
    2.7$, for $L=64$, and $\alpha\approx 2.5$, for $L=128$. For all
    points, the error bars are smaller than the symbols.}
  \label{cond2D}
\end{figure}

Next, we investigate the optimal flow conditions for small-world
networks built over two-dimensional substrates, namely, $L\times L$
square lattices. As displayed in Fig.~\ref{cond2D}, our results show
that the maximum conductance is obtained for $\alpha=0$, in the case
where all conductances are equal, $\delta=0$ (main graph), and for
$\alpha\approx 2$, in the case where the local conductance is
inversely proportional to the link length, $\delta=1$ (inset). These
values give support to the conjectures that $\alpha_{opt}=0$ for
$\delta=0$, and $\alpha_{opt}=d$ for $\delta=1$.

\begin{figure}[t]
  \includegraphics*[width=8.3cm]{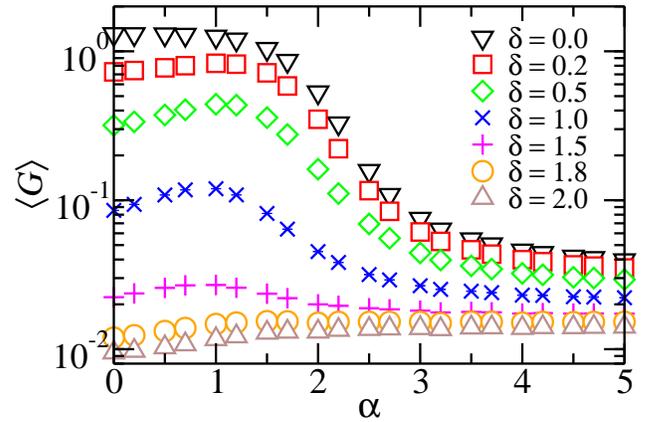}
  \caption{(Color online) Dependence on $\alpha$ of the average global
    conductance of the network for different values of $\delta$. The
    underlying substrate is a one-dimensional regular lattice with
    $N=2048$. The optimal global conductance is found for three
    different regimes of $\delta$: for $\delta=0$, $\alpha_{opt}=0$;
    for $0.2\leq \delta<1.8$, $\alpha_{opt}=1$; and for $\delta\geq
    1.8$, the conductances always grow with $\alpha$, namely, the
    optimal condition is $\alpha\rightarrow \infty$.}
  \label{cond_delta}
\end{figure}

To explore more deeply the phenomenon beyond the Pouillet's law
($\delta=1$), we now study in detail the dependence of the global
conductance on the parameter $\delta$. Note that, in the absence of
long-range connections, the effective conductance between any pair of
sites grows linearly with the distance $r$. That is, for $\delta>1$,
these effective conductances are always larger than the conductances
of their long-range connections in parallel. In this case, therefore,
long-range connections do not play the role of shortcuts. For
$\delta>0$, the longer the added connections, stronger should be their
effect on the global conductance. On the other hand, the longer the
connections, more resistive they are, with less impact on the
effective conductance. These competing effects are responsible for the
observed non-monotonicity of the average conductance with the exponent
$\alpha$. Such a conclusion is supported by the results shown in
Fig.~\ref{cond_delta}, where an optimal $\alpha$ is observed for
$0.2\le\delta<1.8$. We cannot, however, exclude the possibility that
finite size effects are hiding the optimum condition in $\delta=1.8$.

Our results present an striking connection with the problem of
navigation in small-world networks. There we have $\alpha_{opt}=0$ for
global knowledge~\cite{Kosmidis2008} and $\alpha_{opt}=d$ for local
knowledge~\cite{Kleinberg2000}. Here we obtain the same optimal
conditions for equal link conductances and for conductances that
decrease with the link length, respectively. In a sense, solving
Kirchhoff's laws involves more global knowledge than finding the
minimum path, since the flux balances are susceptible to small
disturbances, like the addition or removal of a single conducting link
anywhere in the lattice.  Therefore, it is somewhat surprising that we
obtain for $\delta=1$ the same optimal condition as in the case of
navigation with local knowledge. The key feature of this result is the
way the parameter $\alpha$ controls the length of long-range
connections. If $\alpha$ is small, long connections become frequent.
However, since $\delta>0$, their associated conductances are low. On
the other hand, if $\alpha$ is too large, the conductances of added
connections do not vanish, but their lengths are too small to impact
the scaling.

In conclusion, our results showed in what conditions enhanced flow can
be observed in small-world networks. Long-range interactions are known
to strongly affect the physical properties of real systems. Often the
amount of these interactions is parametrized either by considering a
vanishing strength for the interaction or a vanishing probability for
establishing it. Here we considered the combined effect of these two
conditions by associating power laws for both ({\it i}) the
probability distribution of distances for long-range links,
$P_{ij}\sim r_{ij}^{-\alpha}$, and ({\it ii}) their corresponding
conductances, $g_{ij}\sim{r_{ij}^{-\delta}}$. For $\delta=0$, the
longer the link, the stronger its impact on the flow, leading to
$\alpha_{opt}=0$.  For $\delta>2$, longer random links have
decreasingly small conductances.  In this regime, increasing the
probability of longer connections is detrimental to the conductivity,
and we obtained $\alpha_{opt}\to\infty$. For intermediate values,
$0<\delta\ll{2}$, we observed an optimal condition at
$\alpha_{opt}=d$.  Interestingly, the same optimal conditions were
verified for the problems of navigation with global
knowledge~\cite{Kosmidis2008}, $\alpha_{opt}=0$, and local
knowledge~\cite{Kleinberg2000}, $\alpha_{opt}=d$.  Moreover, in the
case of Pouillet's law, $\delta=1$, we observed that, in the optimal
condition, the conductance vanishes slowly with the size of the system
$\langle G\rangle\propto (\log_{10}N)^{-\gamma}$, with $\gamma=2.17$
for one-dimensional substrates.

We thank the Brazilian Agencies CNPq, CAPES, FUNCAP and FINEP, the
FUNCAP/CNPq Pronex grant, and the National Institute of Science and
Technology for Complex Systems in Brazil for financial support.


\begin{thebibliography}{00}
\bibitem{notsolong} Although we use the term long range generically,
  the length of the added connection may be small, varying from 2 to
  N/2.
\bibitem{footnote} Due to the large amount of stored data, which can
  easily reach hundreds of gigabytes, we average the conductance over
  10\% of all possible pairs of sites in the network, randomly chosen
  in each realization.
\end{thebibliography}
\end{document}